\documentclass[prb,aps,twocolumn,floatfix,showpacs,letter]{revtex4}
\usepackage{color}
\usepackage{bm}
\usepackage{graphicx}

\begin{document}
\title{Theoretical analysis of the transmission phase shift of a
quantum dot in the presence of Kondo correlations}

\author{A. Jerez$^{1,2}$, P. Vitushinsky$^{3}$, and M. Lavagna$^{3}$$^{,*}$}
\affiliation{$^1$European Synchrotron Radiation Facility, 6, rue
Jules Horowitz, 38043 Grenoble Cedex 9, France}
\affiliation{$^2$Institut Laue Langevin, 6, rue Jules Horowitz,
38042 Grenoble Cedex 9, France}
\affiliation{$^3$Commissariat \`a
l'Energie Atomique, DRFMC/SPSMS, 17, rue des Martyrs, 38054
Grenoble Cedex 9, France}

\begin{abstract}
We study the effects of Kondo correlations on the transmission
phase shift of a quantum dot coupled to two leads in comparison
with the experimental determinations made by Aharonov-Bohm (AB)
quantum interferometry. We propose here a theoretical
interpretation of these results based on scattering theory
combined with Bethe ansatz calculations. We show that there is a
factor of 2 difference between the phase of the S-matrix
responsible for the shift in the AB oscillations, and the one
controlling the conductance. Quantitative agreement is obtained
with experimental results for two different values of the coupling
to the leads.

\end{abstract}

\maketitle Quantum dots (QD), small puddles of electrons connected
to leads, can be obtained in a controlled fashion thanks to recent
progress in nanolithography. Under certain conditions a dot can be
modeled as a localized spin coupled to Fermi baths (the leads). A
Kondo effect takes place
\cite{goldhaber}$^,$\cite{cronenwett}$^,$\cite{vanderwiel} when
the temperature is lowered . A key ingredient of the Kondo effect
is the phase shift $\delta$ an electron undergoes when it crosses
the dot. While its direct measurement was out of scope in bulk
systems, it became feasible recently in quantum dots via
Aharonov-Bohm (AB) interferometery\cite{heiblum1}. We mention here
the experimental results obtained in two cases corresponding to a
strong coupling to the leads\cite{heiblum1}$^,$\cite{heiblum2}. In
the unitary limit, the phase shift climbs almost linearly with
$V_G$ with a value at the middle of the Kondo enhanced valley
which is almost $\pi$. At a smaller value of the coupling
strength, the phase shift develops a wide plateau at almost $\pi$.
We will call the latter case the "Kondo regime". Early theoretical
work on the phase shift for the bulk Kondo effect
\cite{langreth}$^,$\cite{nozieres} predicts $\delta=\pi/2$. In the
context of QD, Gerland et al\cite{gerland} had obtained, on the
basis of NRG and Bethe ansatz calculations, a variation of
$\delta$ with the energy of the localized state leading to a value
of $\pi/2$ in the symmetric limit, in disagreement with the recent
experimental results quoted
above\cite{heiblum1}$^,$\cite{heiblum2}. In this paper, we propose
a new theoretical interpretation of the experimental results based
on scattering theory and Bethe ansatz calculations. Our main
prediction concerns a factor of 2 difference found between the
phase of the S-matrix observed by the phase shift measurements and
the phase governing the conductance.

Let us consider a quantum dot coupled via tunnel barriers to two
leads $L$ and $R$, and capacitively to a gate maintained at the
voltage $V_G$. The system can be
described\cite{glazmanraikh}$^,$\cite{nglee} by an one-dimensional
Anderson model with two reservoirs $L$ and $R$
\begin{eqnarray}\label{anderson}
\nonumber
H=-t\sum_{\sigma}[\sum_{i\geq{1}}(c^{\dag}_{i,\sigma}c_{i+1,\sigma}+h.c.)
+\sum_{i\leq{-2}}(c^{\dag}_{i,\sigma}c_{i+1,\sigma}+h.c.)]\\
\nonumber
-V_{R}\sum_{\sigma}(c^{\dag}_{0,\sigma}c_{1,\sigma}+h.c.)
-V_{L}\sum_{\sigma}(c^{\dag}_{-1,\sigma}c_{0,\sigma}+h.c.)\\
+ \varepsilon_0\sum_{\sigma} n_{0\sigma} +
Un_{0\uparrow}n_{0\downarrow},
\end{eqnarray}
Consider the elastic component of the S-matrix,
$\hat{S}_{k\sigma}$, describing the scattering of a spin-$\sigma$
electron with momentum k off the impurity. It is given
by\cite{merzbacher}$^,$\cite{langreth}$^,$\cite{nglee}
$\hat{S}_{k\sigma}=\textrm{C}_{\sigma}(\hat{I}-i\hat{T}_{k\sigma}^{res})$,
where $\textrm{C}_{\sigma}$ is a multiplicative phase factor and
$\hat{T}_{k\sigma}^{res}$ is the T-matrix with matrix elements
given by
\begin{equation}\label{T1}
T^{res,\alpha\beta}_{k\sigma}=2\pi V_{\alpha}
V_{\beta}\rho_{\sigma}(\varepsilon_k)\mathcal{G}_{\sigma}(\varepsilon_k+i\eta),
\end{equation}
\noindent where $\alpha, \beta=L$ or $R$,
$\rho_{\sigma}(\varepsilon_k)$ is the density of states of
conduction electrons for $\sigma$ and $\varepsilon_k$, and
$\mathcal{G}_{\sigma}(\varepsilon_k+i\eta)$ is the exact localized
electron retarded Green's function. Using exact
results\cite{langreth}$^,$\cite{hewson} on the self-energy at
$T=0$ in an interacting Fermi liquid, one can show that
$n_{0\sigma}=\frac{1}{\pi}
Im\ln{\mathcal{G}_{\sigma}(\mu+i\eta)}$. Friedel's sum
rule\cite{friedel}$^,$\cite{hewson} requires $n_{0\sigma}$ to be
equal to the change in the number of conduction electrons with
spin $\sigma$ resulting from the addition of the impurity. Hence
it is related to the transmission phase shift $\delta_{\sigma}$ at
the Fermi level, $n_{0\sigma}=\frac{1}{\pi}\delta_{\sigma}$.
Therefore $\delta_{\sigma}$ coincides with the phase of the
Green's function at the Fermi level
$\mathcal{G}_{\sigma}(\mu+i\eta)$. If we denote the associated
self-energy by $\Sigma_{\sigma}(\mu+i\eta)$, one gets
$\mathcal{G}_{\sigma}(\mu+i\eta)=\sin\delta_{\sigma}e^{i\delta_{\sigma}}/Im\Sigma_{\sigma}(\mu+i\eta)$.
with
$Im\Sigma_{\sigma}(\mu+i\eta)=-\pi(V_{L}^{2}+V_{R}^{2})\rho_{\sigma}(\mu)$
\cite{langreth}$^,$\cite{hewson} at $T=0$ leading to
\begin{equation}\label{T2}
    T^{res,\alpha\beta}_{k_F\sigma}=-2\frac{V_{\alpha}V_{\beta}}{(V_{L}^{2}+V_{R}^{2})}\sin{\delta_{\sigma}}e^{i\delta_{\sigma}}.
\end{equation}
In the case of a symmetric QD with $V_{L}=V_{R}=V$, one has
$S^{LR}_{k_F\sigma}=S^{RL}_{k_F\sigma}=\textrm{C}_{\sigma}i\sin{\delta_{\sigma}}e^{i\delta_{\sigma}}$
and
$S^{LL}_{k_F\sigma}=S^{RR}_{k_F\sigma}=\textrm{C}_{\sigma}\cos{\delta_{\sigma}}e^{i\delta_{\sigma}}$.
The multiplicative phase factor $\textrm{C}_{\sigma}$ contains
additional information about the spectrum and the filling of the
quantum dot. To determine it, we will make use of Levinson's
theorem\cite{levinson}$^,$\cite{schiff}. In its original form, the
theorem applies to the potential scattering of a particle in a
given momentum $l$ and relates the zero-energy phase shift
$\delta_{l}$ to the number of bound states of the same $l$
supported by the potential. It was
generalized\cite{rosenberg1}$^,$\cite{rosenberg2} later on to the
case of the scattering of a particle by a neutral compound system
as constituted for instance by an atom. In the present case of a
QD which can be viewed as an artificial atom, it follows that
$\ln\det\hat{S}_{k_F\sigma}/(2i\pi)$ is equal to the total number
of states i.e. $\sum_{\sigma}n_{0\sigma}=n_{0}$. By applying
generalized Levinson's theorem to $\hat{S}_{k_F\sigma}$, one finds
$\textrm{C}_{\sigma}=e^{i\delta_{-\sigma}}$ and
\begin{equation}\label{S}
    \hat{S}_{k_F\sigma}=e^{i\delta}
    \left(
    \begin{array}{cc}
      \cos{\delta_{\sigma}} & i\sin{\delta_{\sigma}} \\
      i\sin{\delta_{\sigma}} & \cos{\delta_{\sigma}} \\
    \end{array}
    \right),
\end{equation}
\noindent where $\delta=\sum_{\sigma}\delta_{\sigma}$. One can
easily check that, $\hat{S}_{k_F\sigma}$ being a unitary matrix,
the optical theorem is fulfilled:
$\hat{T}_{k_F\sigma}\hat{T}^{\dag}_{k_F\sigma}=-2Im\hat{T}_{k_F\sigma}$,
where $\hat{T}_{k_F\sigma}=-i(1-\hat{S}_{k_F\sigma})$.
\begin{figure}[h]
    \centering
    \includegraphics[width=0.70\columnwidth]{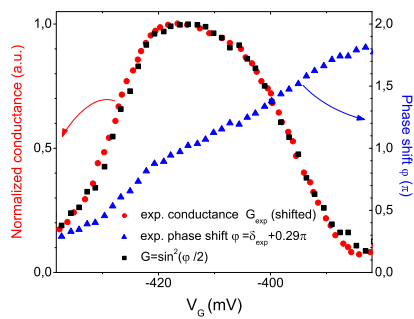}
    \caption{Experimental conductance $G_{exp}(V_{G})$
and phase shift $\varphi(V_{G})$ as a function of $V_G$ (values
taken from Ref.\onlinecite{heiblum1} cf. text). Comparison is made
with the curve $G(V_{G})=\sin^{2}(\varphi(V_{G})/2)$.}
\label{FIG1}
\end{figure}

In an open Aharonov-Bohm interferometry experiment\cite{heiblum2},
spin-$\sigma$ electrons coming from the source through each of the
two arms interfere coherently  at the drain, leading to periodic
oscillations of the differential conductance, the argument of
which is given by $2\pi\Phi e/h+\delta_{QD}$. $\Phi$ is the
magnetic flux and $\delta_{QD}$ is the transmission phase shift
introduced by the QD, equal to $\delta=\pi n_{0}$ (cf.
Eq.\ref{S}). In this paper, we neglect the role of the reference
arm on the phase shift considered by some authors\cite{hofstetter}
and concentrate on the contribution of the quantum dot to the
interference pattern. The conductance through the QD is expressed
by the Landauer formula\cite{landauer}$^,$\cite{meir},
$G\propto\sum_{\sigma}|T^{LR}_{k_{F}\sigma}|^{2}$. Using
Eq.\ref{S}, we get
$G\propto\sum_{\sigma}\sin^{2}{\delta_{\sigma}}$. In the absence
of magnetic field,
$\delta_{\uparrow}=\delta_{\downarrow}=\delta/2$, one gets
\begin{equation}\label{G}
    G\propto\sum_{\sigma}\sin^{2}{\delta/2}.
\end{equation}
\begin{figure}
    \centering
    \includegraphics[width=0.60\columnwidth]{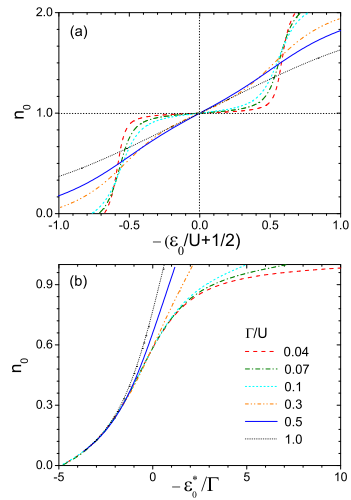}
    \caption{(a) Theoretical results from the B.A. calculations
for the occupation number $n_0$ as a function of the normalized
energy $(-\varepsilon_{0}/U+1/2)$ at different values of
$\Gamma/U$. Note that $n_{0}=1$ at the symmetric limit
$\varepsilon_{0}=-U/2$ and the existence of a plateau in its
vicinity when $\Gamma/U \leq 0.25$; (b) The same quantity as a
function of the renormalized energy $\varepsilon^*/\Gamma$ at
different values of $\Gamma/U$.} \label{FIG2}
\end{figure}

Due to recent developments in experimental techniques, one now
disposes of simultaneous measurements of $G$ and $\delta_{QD}$. In
this paper we check the validity of the theoretical prediction of
Eqs.(\ref{S},\ref{G}) by reporting the experimental results for
$G$ and $\delta_{QD}$ obtained in the unitary limit at different
values of $V_G$. Before examining the experimental test, we make
the following remarks: (i) in an interferometry experiment, only
relative values of the transmission phase shifts can be measured.
Hence we set $\delta=\pi$ at the location of the maximum of the
visibility, evaluated to $V_{G}=423mV$. This implies a shift in
the $\delta$-scale evaluated to $\Delta\delta=0.29\pi$ with
$\varphi=\delta+\Delta\delta$; (ii) typically the measurement of
the conductance $G$ is performed in a "one-arm" device (pinching
off the reference arm with the barrier gate) whereas that of the
visibility is done in a "two-arm" device. As a result, while the
evolution of the visibility with $V_{G}$ mimics that of the
conductance, the value of the former is shifted with respect to
that of the latter, by $\Delta V_G=15mV$. Therefore we take the
values of $G$ at $(V_G-\Delta V_G)$, and of $\delta$ at $V_G$;
(iii) the conductance is normalized by its maximum value at
$V_{G}=423mV$. Taking all these points into account, the graph
reported in Fig.\ref{FIG1} shows that the experimental dependence
of $\sin^{2}{\varphi/2}$ with $V_{G}$ reproduces that of the
"shifted" conductance $G_{exp}$ in a quite remarkable way,
providing further support to the validity of
Eqs.(\ref{S},\ref{G})\cite{note1}.

We now want to evaluate $n_0$ in order to derive $\delta=\pi n_0$.
Starting from Eq.(\ref{anderson}), one can show\cite{glazmanraikh}
that only the symmetric linear combination of electrons couples to
the localized state. Therefore if we are only interested in
$n_{0}$, it is sufficient to study a single reservoir Anderson
model with a hybridization potential
$\widetilde{V}=\sqrt{V_{L}^{2}+V_{R}^{2}}$. We have solved the
equations of the Bethe ansatz (B.A.) numerically at
$T=0$\cite{kawakami1}$^,$\cite{kawakami2}$^,$\cite{tsvelick}. This
allows us to determine the value of $n_0$ as a function of the
parameters of the Anderson model $\varepsilon_0$, $V$ and $U$. The
three parameters enter through their ratios $\varepsilon_0/U$ and
$\Gamma/U$, where $\Gamma=\pi V^2 \rho_0$. Denote by
$n_0(\varepsilon_0,\Gamma/U)$ the value of $n_0$ for the
corresponding values of the parameters. The following relation
holds due to the particle-hole symmetry of the
model\cite{kawakami2}:
$n_0(-(\varepsilon_0+U),\Gamma/U)=2-n_0(\varepsilon_0,\Gamma/U)$.
This automatically ensures $n_0(-U/2,\Gamma/U)=1$ in the symmetric
limit $\varepsilon_0=-U/2$ whatever $\Gamma/U$ is. Furthermore it
follows from the preceding relation that the study can be
restricted to $-U/2\leq\varepsilon_0\leq U/2$ and the remaining
part can be deduced from it. The results of the calculations are
reported in Fig.2(a). For strong coupling strengths $\Gamma/U \geq
0.25$, $n_0$ is found to climb almost linearly with
$-(\varepsilon_0/U+1/2)$ whereas for weak coupling strengths
$\Gamma/U\leq 0.25$, the energy dependence of $n_0$ develops a
plateau around $\varepsilon_0=-U/2$. This change of behavior is
due to the fact that the extent of the local moment regime
(centered around $\epsilon_0=-U/2$ with $n_0\simeq 1$) increases
when $\Gamma/U$ decreases. As the temperature is lowered, the
Kondo resonance develops through this local moment regime. This
plateau-like structure can be viewed as the beginnings of the
"staircase" variation of $n_0$ with $\varepsilon_0$ obtained in
the localized regime $\Gamma/U\rightarrow0$.

\begin{figure}
    \centering
    \includegraphics[width=0.54\columnwidth]{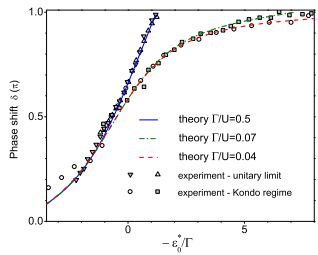}
    \caption{Fit of the experimental data for the $V_{G}$-dependence
of the phase shift with the B.A. results for $\delta=\pi n_{0}$ as
a function of $-\varepsilon_0^{*}/\Gamma$. Making use of the
electron-hole symmetry, experimental results both below and above
the symmetric limit are reported using the same scales. They are
represented by triangles pointing down and up respectively in the
unitary limit, and by circles and squares respectively in the
Kondo regime (incorporating a shift in the $\delta$-scale
cf.text). The best fit is obtained for $\Gamma/U=0.5$ in the
unitary limit (both below and above the symmetric limit), and for
$\Gamma/U=0.07$ or $0.05$ in the Kondo regime (below or above the
symmetric limit respectively).} \label{FIG3}
\end{figure}

\begin{figure}[t]
    \centering
    \includegraphics[width=0.50\columnwidth]{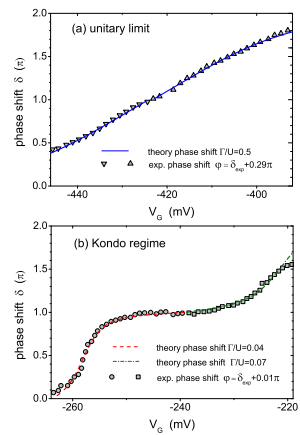}
    \caption{Phase shift as a function of the gate voltage $V_{G}$.
(a) Unitary limit. Theoretical results from Bethe ansatz
calculations at $\Gamma/U=0.5$ compared to the experimental data
for $\varphi/\pi=\delta_{exp}/\pi+0.29$ (triangles pointing down
and up) (b) Kondo regime. Same thing with $\Gamma/U=0.04$ and
$0.07$, $\varphi/\pi=\delta_{exp}/\pi+0.01$ (circles and
squares).} \label{FIG4}
\end{figure}

The experimental data can be fitted then with two parameters,
$\Gamma/U$ and $\varepsilon_0/U$. The value of $\varepsilon_0$ is
governed by the strength of the gate voltage. Fitting the
experimental data from results presented in Fig.2(a) is a
difficult task since one needs to fix the correspondence between
$\varepsilon_0/U$ and $V_G$ on the one hand (we take it linear as
usual, independent of the regime considered), and to find the best
fitting value for $\Gamma/U$ in the different regimes on the other
hand. A valuable help for doing this is provided by taking
advantage of some special properties of the Anderson model. These
properties can be easily recognized when physical quantities such
as $n_0$ are plotted as a function of some renormalized energy
defined as $\varepsilon^*/\Gamma = \varepsilon_0/\Gamma +
g(U/(\Gamma))$. In the asymmetric regime\cite{tsvelick} when
$(U+2\varepsilon_0)\gg \sqrt{\Gamma U}$ and $|\varepsilon_0|\ll
U$, $g(U/(\Gamma)$ equals $\frac{1}{\pi}\ln(\pi e U/(4\Gamma))$
and the behavior of $n_0$ as a function of
$-\varepsilon_0^{*}/\Gamma$ is
universal\cite{haldane}$^,$\cite{tsvelick}. This property is
illustrated in Fig.2(b). The universality is reached when
$\Gamma/U \leq 0.25$ and the range of energy over which universal
behavior extends is given by
$|\varepsilon_0^{*}/\Gamma-1/\pi\ln(\alpha U/\Gamma)| \ll
U/\Gamma$. One can also see from Fig.2(b) that in the empty level
regime ($n_0\rightarrow0$), the curves
$n_0=f(\varepsilon_0^{*}/\Gamma,\Gamma/U)$ at various values of
$\Gamma/U$ merge, displaying an asymptotic behavior\cite{note3}.
The existence of both these universal and asymptotic behaviors is
of valuable help in fitting the experimental data. Fig.\ref{FIG3}
reports the results of the fit in the unitary limit and Kondo
regimes. The experimental results incorporate a shift in the
$\delta$-scale, $\varphi=\delta+\Delta\delta$ in order to get
$\varphi=\pi$ at the symmetric limit. We establish the
correspondence between $V_{G}$ and $\varepsilon_{0}/U$ by fitting
the experimental data to the theoretical results in the empty
level regime when all the curves merge. One finds $\Delta
V_{G}/\Delta(\varepsilon_{0}/U)$ of the order of $30mV$ in both of
the regimes considered. The best fit is obtained for
$\Gamma/U=0.5$ in the unitary limit both below and above the
symmetric limit, and for $\Gamma/U=0.04$ or $0.07$ in the Kondo
regime (below or above the symmetric limit respectively). Finally
by keeping the same correspondence between $V_{G}$ and
$\varepsilon_{0}/U$ and using $\delta=\pi n_{0}$, we derive the
dependence of the phase shift with $V_{G}$ from results obtained
in Fig.2(a). As can been seen from Fig.\ref{FIG4}, our theoretical
predictions are in quantitative agreement with the experimental
data. The fit is all the more remarkable that it is performed in
presence of a single fitting parameter $\Gamma/U$ only.

In conclusion, we have proposed a theoretical analysis of the
transmission phase shift of a quantum dot in presence of Kondo
correlations and confronted our results with the Aharonov-Bohm
interferometry and conductance measurements. We have explained the
presence of a factor of 2 difference between the total phase of
the S-matrix (responsible for the shift in the A-B oscillations),
and the one appearing in the expression of the conductance
$G\sim\sin^{2}(\delta/2)$. Our calculations based on Bethe ansatz
lead to a remarkable quantitative agreement with experimental
results. The whole discussion so far has been restricted to the
low temperature regime. One of the next goals will be to include
finite temperature effects as well as to study the role of a
magnetic field and consider the out-of-equilibrium situation.

We would like to thank N. Andrei for his encouragements and
discussions during the course of this work. We would also like to
thank P. Nozi\`{e}res and E. Kats for helpful discussions.  A.J.
thanks the hospitality of the Physics Department at Rutgers
University. M.L. acknowledges the hospitality of the Aspen Center
for Physics and the Kavli Institute for Theoretical Physics in
Santa Barbara where part of this work has been done. This work was
supported by the Institute for Condensed Matter Physics (IPMC) of
Grenoble.

* Also at the Centre National de la Recherche Scientifique (CNRS).


\begin{thebibliography}{99}
\bibitem{note1}
Note that Eq.(\ref{G}) is only valid when $T\ll T_K$. Whereas the
condition is satisfied in the experiments performed in the unitary
limit, the relation does not hold for experiments performed in the
Kondo regime when weaker coupling strength introduces a reduction
of $T_K$.
\bibitem{note3}
This can be proved by noticing that $\Gamma$ is almost
unrenormalized in the $n_0\rightarrow 0$ limit.
\end{thebibliography}
\end{document}